# Optical Clocks at Sea

Jonathan D. Roslund*, Arman Cingöz, William D. Lunden, Guthrie B. Partridge, Abijith S. Kowligy, Frank Roller, Daniel B. Sheredy, Gunnar E. Skulason, Joe P. Song, Jamil R. Abo-Shaeer, & Martin M. Boyd*

Vector Atomic, Inc., Pleasanton, CA, USA

## Abstract
Deployed optical clocks will improve positioning for navigational autonomy [1], provide remote time standards for geophysical monitoring [2] and distributed coherent sensing [3], allow time synchronization of remote quantum networks [4] [5], and provide operational redundancy for national time standards. While laboratory optical clocks now reach timing inaccuracies below $10^{-18}$ [6] [7], transportable versions of these high-performing clocks [8] [9] have limited utility due to their size, environmental sensitivity, and cost [10]. Here we report the development of optical clocks with the requisite combination of size, performance, and environmental insensitivity for operation on mobile platforms. The 35 L clock combines a molecular iodine spectrometer, fiber frequency comb, and control electronics. Three of these clocks operated continuously aboard a naval ship in the Pacific Ocean for 20 days while accruing timing errors below 300 ps per day. The clocks have comparable performance to active hydrogen masers in one-tenth the volume. Operating high-performance clocks at sea has been historically challenging and continues to be critical for navigation. This demonstration marks a significant technological advancement that heralds the arrival of future optical timekeeping networks.

## Introduction

Atomic timekeeping plays an essential role in modern infrastructure, from transportation to telecommunications to cloud computing. Billions of devices rely on the Global Navigation Satellite System (GNSS) for accurate positioning and synchronization [11]. GNSS is a network of distributed, high-performance microwave-based atomic clocks that provide nanosecond-level synchronization globally. The emergence of fieldable optical timekeeping, which offers femtosecond (fs) timing jitter at short timescales and multiday, sub-nanosecond holdover, along with long distance, fs-level optical time-transfer [12], paves the way for global synchronization at picosecond levels.

Molecular iodine ($I_2$) has a legacy as an optical frequency standard [13] [14] [15] [16] [17]. Several iodine transitions are officially recognized as length standards [18], and the species underpinned one of the first demonstrations of optical clocks [19]. More recently, iodine frequency standards have been investigated for space missions [20] [21] [22] [23]. Here we report the deployment of multiple high-performance, fully integrated iodine optical clocks and highlight their ability to maintain nanosecond- (ns) level timing for several days while continuously operating at sea.

* Corresponding authors: jon@vectoratomic.com, marty@vectoratomic.com

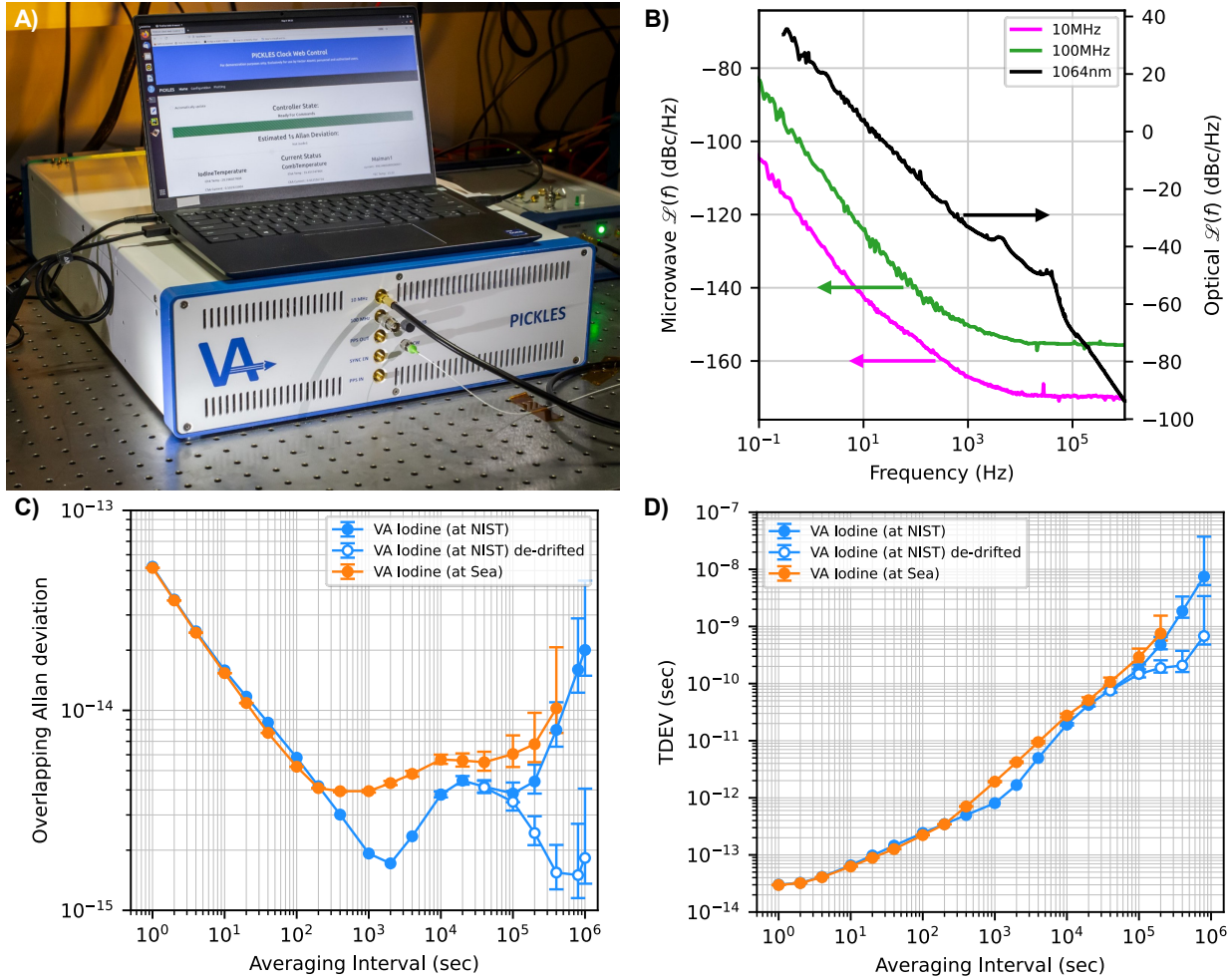

Figure 1. Single clock performance at NIST and at sea. A) The 3U, 19-inch rackmount iodine optical clock occupies a volume of 35 L and consumes less than 100 W. B) Measured phase noise for the iodine clock at 10 MHz, 100 MHz, and 1064 nm. C) Overlapping Allan deviation for the iodine clock operating at NIST and at-sea. At short timescales, the instability in a dynamic environment is identical to the laboratory. The iodine clock can maintain < $10^{-14}$ frequency instability for multiple days despite several degree temperature swings, significant changes in relative humidity, and changing magnetic fields. D) The clocks can maintain holdovers of 10 ps for several hours and 1 ns for multiple days, showing their potential as the basis for a picosecond-level timing network.

These clocks employ a robust vapor cell architecture that uses no consumables, does not require laser cooling or a pre-stabilization cavity, and is first-order insensitive to platform motion. Similar approaches with rubidium vapor cells are under development [24] [25] [26]. Importantly, iodine clocks employ mature laser components at 1064 nm and 1550 nm. The focus on a robust laser system rather than a high-performance atomic species resolves system-level issues with dynamics, lifetime, autonomy, and cost. While not as accurate as laboratory optical clocks using trapped atoms or ions, iodine clocks can provide maser-level performance in a compact, robust, and mobile package.

Initial clock prototypes, shown in Figure 1A, were integrated into 35 L, 3U 19-inch rackmount chassis. Clock outputs are at 100 MHz, 10 MHz, and 1 pulse per second (PPS). Auxiliary optical outputs are provided for the frequency comb and clock laser (1550 nm and 1064 nm, respectively). The physics packages, which include the spectrometer, laser system, and frequency comb were designed and built in-house to reduce system-level size, weight, and power (SWaP). The integrated control system is a mix of custom and commercial circuit boards, with custom firmware. The clock operates using a commercial 1U rackmount power supply and control laptop.

Two identical clocks (PICKLES and EPIC) were developed with physics packages targeting short-term instability below $10^{-13}/\sqrt{\tau}$. A third clock (VIPER) was built using a smaller iodine spectrometer and laser system to reduce physics package SWaP at relaxed performance goals. The frequency comb design and control electronics for PICKLES, EPIC, and VIPER are largely identical.

**Results**

In April 2022, PICKLES and EPIC were shipped to the National Institute of Standards and Technology (NIST) in Boulder, Colorado for assessment against the UTC(NIST) timescale [27]. The clocks operated on an optical table without any special environmental measures and the laboratory was in active use throughout the measurement campaign. The 10 MHz tone from each clock was compared against a 5 MHz maser signal with a Microchip 53100A phase noise analyzer in a three-cornered hat configuration. NIST maser ST05 (Symmetricom MHM-2010) was selected as the lowest drift maser in the ensemble ($3 \times 10^{-17}$/day). The measurement scheme allows for decorrelating the three clocks at short timescales and measuring against the NIST composite timescale AT1, derived from the maser ensemble, at longer timescales. Importantly, ST05 was operated in an environmental chamber in a separate laboratory, providing an environmentally uncorrelated reference. The 1064-nm optical beatnote between the PICKLES and EPIC was simultaneously monitored for cross-validation. After installation, the clocks were left to operate autonomously. The measurement set-up was remotely monitored without intervention from our California headquarters, and the comparison was intentionally terminated after 34 days on return to NIST.

The overlapping Allan deviation for the entire 34-day dataset without any windowing, de-drifting, or filtering is shown in Figure 2. To present the individual clock performance, the Allan deviation plot uses the 1-1,000 s instability extracted from three-cornered hat analysis and the direct instability against ST05 for time periods longer than 1,000 s (see supplemental material for raw measurements). The PICKLES and EPIC short-term instabilities of $5 \times 10^{-14}/\sqrt{\tau}$ and $6 \times 10^{-14}/\sqrt{\tau}$, respectively, outperform the short-term performance of the ST05 maser. Both iodine clocks exhibit fractional frequency instabilities $< 5 \times 10^{-15}$ after 100,000 s of averaging, equivalent to a temporal holdover below 300 ps after 1 day.

The data also provided an initial measure of the long-term stability of the iodine clocks (Figure 2, inset). Measured against UTC(NIST), the drift rates for PICKLES and EPIC are $\sim 2 \times 10^{-15}$ and $\sim 4 \times 10^{-15}$/day, respectively, consistent with the long-term accuracy of an iodine vapor cell measured over the course of a year [19]. Moreover, the observed drift rate is $\sim 10 \times$ lower than a typical space-qualified rubidium atomic frequency standard (RAFS) after more than a year of continuous operation [28] [29]. Removal of a constant linear drift from the frequency data indicates that the two clocks continue to hold $< 3\times10^{-15}$ instability after more than $10^6$ seconds (~12 days) of averaging, equivalent to 1 ns timing error over this period. Without drift removal, the long-term clock performance is competitive with active hydrogen masers; drift removal puts the clock stability on-par with the highest performing masers. Notably, to achieve the drift rates observed in Figure 2, the NIST masers are operated continuously for years and housed in ~1,000 L environmental chambers to stabilize temperature and humidity to better than 100 mK and 1%, respectively [30] [31]. The laboratory housing PICKLES and EPIC was stable to 100's of mK throughout the measurement campaign, which started a few days after a cross country shipment. Finally, the raw iodine clock performance is below NIST's commercial cesium beam clock (Microchip 5071A) for ~5.5 days; the de-drifted iodine performance is below cesium for all observed timescales.

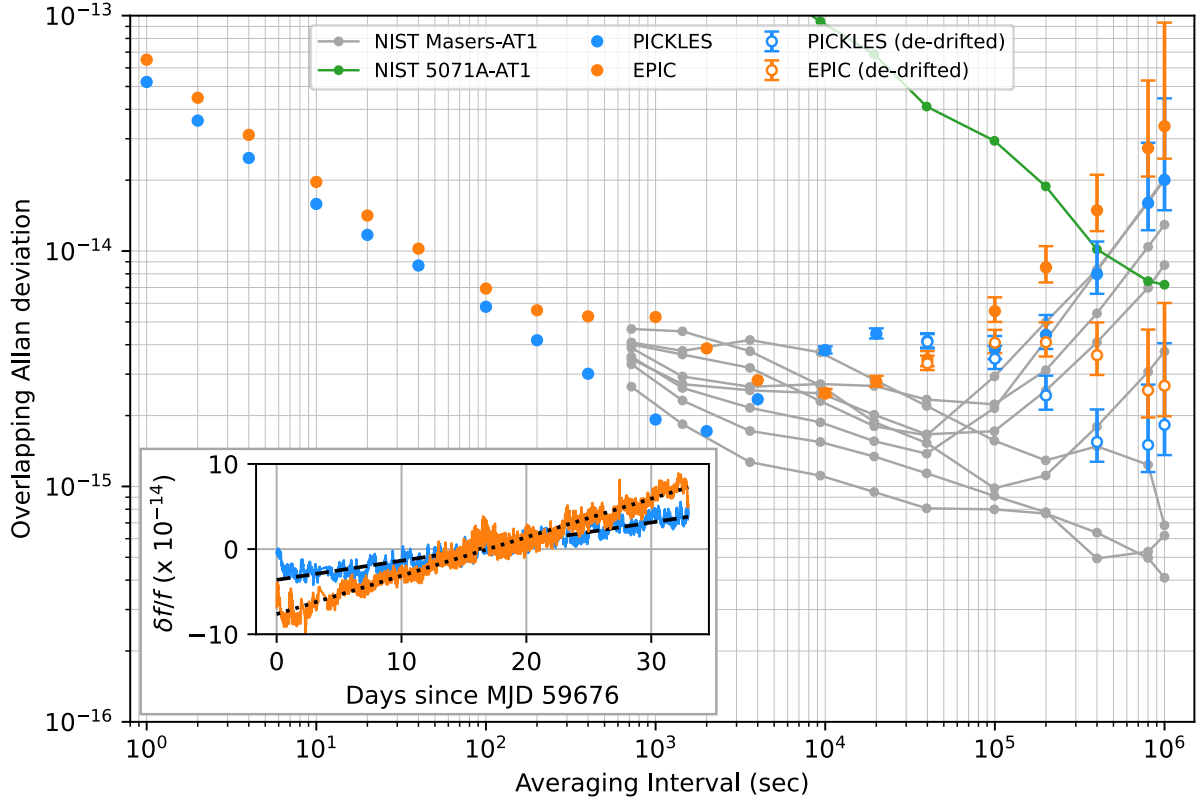

Figure 2. Long-term clock performance. Overlapping Allan deviation for the 10 MHz outputs of the two Iodine clocks measured against the UTC(NIST) timebase for 34 days (blue and orange traces). The clocks exhibit a raw frequency instability of $4\times10^{-15}$ (PICKLES) and $6\times10^{-15}$ (EPIC) after $10^5$ s of averaging and maintain instability $<10^{-14}$ for nearly 6 days (PICKLES). With linear drift removal, the frequency instability improves to $<2\times10^{-15}$ (PICKLES) and $<3\times10^{-15}$ (EPIC) for $10^6$ seconds (open circles). The performance of a variety of NIST masers against the composite AT1 time scale is shown for comparison (gray traces) as well as a commercial cesium clock (green trace). The long-term frequency record for the two iodine clocks against ST05 is shown as an inset. Each trace is shown as a 1,000 second moving average. The linear drift for each clock is observed to be several $10^{-15}$ per day.

A broad feature with a peak deviation of ~$4\times10^{-15}$ is evident in the PICKLES Allan deviation on ~7-hour timescales. The equivalent ~2 Hz optical frequency deviation corresponds to a ~2 ppm shift of the hyperfine transition linewidth. We believe that the origin of this plateau in PICKLES is residual amplitude modulation (RAM) coupling through an uncontrolled etalon in the spectrometer. This etalon was removed during the building of the EPIC spectrometer, and measures have been taken to reduce it in future versions of the clock.

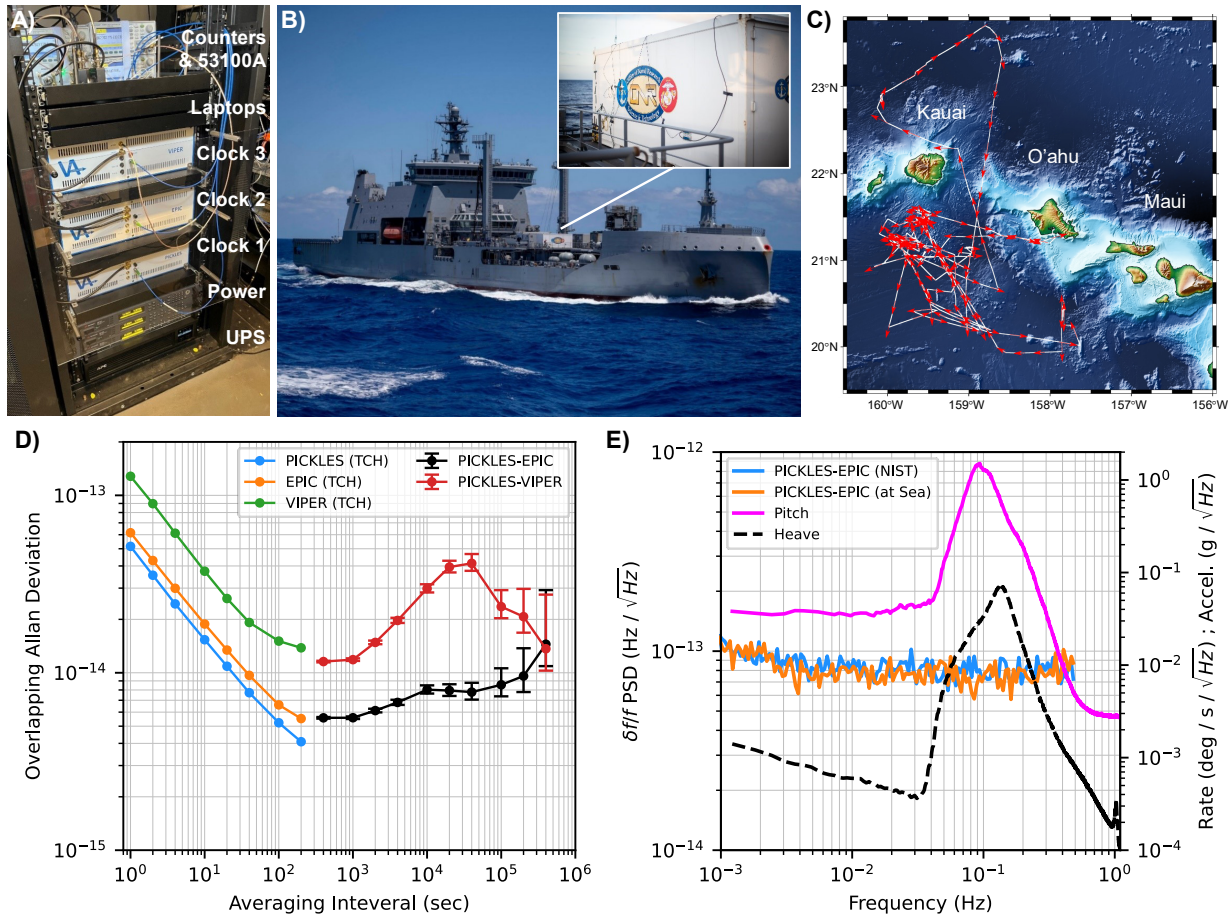

Figure 3. At-sea demonstration of optical clocks. A) Clock stack-up for RIMPAC 2022. The server rack contained three independent optical clocks, a 1 U power supply and control laptop for each clock, an uninterruptable power supply (UPS), and the measurement system in a total rack volume of 23U. B) The cargo container housing the clocks was craned onto the deck of the HMNZS Aotearoa where it remained for the three-week naval exercise [47]. C) A GPS track of the Aotearoa's voyage around the Hawaiian Islands. The ship started and ended its voyage at Pearl Harbor, O'ahu. D) Overlapping Allan deviation during the underway. For time periods less than 100 s, individual clock contributions are extracted with a three-cornered hat (TCH) analysis; directly measured pairwise instabilities are shown for periods longer than 100 s. The EPIC-PICKLES pair maintains a fractional frequency instability of $8\times10^{-15}$ after $10^5$ s of averaging, corresponding to a ~400 ps temporal holdover. E) Power spectral density (PSD) for the PICKLES-EPIC frequency fluctuations at NIST and at-sea with the recorded ship pitch and heave (rotation and acceleration on the other ship axes showed similar behavior). The PICKLES-VIPER PSD (not shown) showed a similar immunity to the ship motion.

The iodine clock exhibits excellent phase noise for the 10 MHz and 100 MHz tones derived by optical frequency division as well as the 1064nm optical output (Figure 1B). The phase noise at microwave frequencies is lower than commercial atomic-disciplined oscillators, highlighting the benefits of optical frequency division where the fractional noise of the iodine-stabilized laser is transferred to the RF comb repetition rate. In addition, the iodine-stabilized laser provides ~100,000× lower drift rate as compared to ultralow expansion (ULE) optical cavities [32].

Following the measurement against an absolute reference at NIST-Boulder, three optical clocks were brought to Pearl Harbor, HI in July 2022 to participate in the Alternative PNT (A-PNT) Challenge at RIMPAC 2022, the world's largest international maritime exercise. A-PNT included additional prototype quantum technologies [33] [34] [35]. The clocks were installed in an open server rack along with a commercial 1U power supply for each clock, three control laptops, and an uninterruptable power supply (UPS) backup for the system (Figure 3A). The rack also contained three frequency counters to collect the three pairwise beatnotes and a 53100A phase noise analyzer to compare the 100 MHz tone derived from each clock's frequency comb against the other two in a three-cornered hat configuration. The total stack-up, including three independent clocks, power supplies, computer controls, and metrology systems occupied a rack height of 23U. The server rack was hard-mounted to the floor of a Conex cargo container, which was craned onto the deck of the New Zealand naval ship HMNZS Aotearoa (Figure 3B), where it remained during the three weeks the vessel was at sea. Once the ship left port, the three clocks operated without user intervention for the duration of the exercise, apart from one restart of VIPER due to a software fault in the external power supply.

The operating environment during the ship's underway differed significantly from NIST but the clocks still operated continuously with high performance (Figure 1C and D). Although the Conex was air-conditioned, the internal environment underwent ~3 °C day-night temperature swings. The clock rack was located directly in front of the AC unit, which cycled on and off throughout the day. The clocks also operated continuously through ship motion. The rotational dynamics of the ship included a peak pitch of ± 1.5° at a rate of ± 1.2°/s and a peak roll of ± 6° at a rate of ± 3°/s. Similarly, the maximum surge, sway, and heave accelerations were ± 0.4, ± 1.5, and ± 1.2 m/s$^2$, respectively. A vertical root mean square (RMS) vibration of 0.03 m/s$^2$ (integrated from 1 Hz to 100 Hz) was also experienced. Operation in dynamic environments highlights the robust, high-bandwidth clock readout enabled by a vapor cell.

The vessel traveled in all four cardinal directions during the exercise, illustrated by the GPS-tracked trajectory in Figure 3C. The NOAA geomagnetic model for Earth's magnetic field at this latitude and longitude reveals that the projection of the earth's field on the clocks varied by ± 270 mG throughout the underway [36].

The overlapping Allan deviations measured during the voyage are shown in Figure 3D. For time-periods less than 100 seconds, the individual clock contributions are extracted with a three-cornered hat analysis. Directly measured pairwise instabilities are shown for longer time periods. There was no degradation in the clock SNR despite ship vibration and motion; the short-term performance for the three clocks was identical to that observed at NIST for up to 1,000 s (Figure 1C and D). All three clocks showed immunity to dominant ship motion in the ~ 0.1 Hz band (Figure 3E). A medium timescale instability was driven by the day-night temperature swing in the Conex. Nonetheless, the PICKLES-EPIC clock pair maintains 8×10$^{-15}$ combined instability at 100,000 s without drift correction, equivalent to ~400 ps temporal holdover over 24 hours. EPIC presents a temperature-driven instability in the $10^4 – 10^5$ s range (identified as EPIC based on environmental chamber testing following RIMPAC), but its performance is still within 2× of that seen at NIST. Finally, the drift rate for PICKLES-EPIC over this period was similar to that observed at NIST. This long-term performance illustrates the robustness of iodine-based timekeeping as the clocks experienced diurnal temperature swings of several degrees, platform motion arising from ship dynamics, and constant movement through Earth's magnetic field.

VIPER exhibits a short-term instability of $1.3\times10^{-13}/\sqrt{\tau}$ as well as a diurnal temperature-driven instability that peaks at $4\times10^{-14}$ at ~40,000 s. The VIPER physics package is an earlier design with relaxed performance goals that results in a larger temperature coefficient than the other two clocks. Nonetheless, this system can average over the diurnal temperature fluctuation and maintain an instability of $2.5\times10^{-14}$ after one day of averaging. VIPER showed a similar drift rate to PICKLES and EPIC during the underway. Importantly, the VIPER physics package does not include magnetic shields yet still provides excellent frequency stability despite motion through Earth's magnetic field. Summary data for the highest performing clock at NIST and at sea is shown in Figure 1C and D.

## Conclusion

Iodine has proven to be a capable platform for the development of practical optical timekeeping systems. The unique combination of SWaP, phase noise, frequency instability, low environmental sensitivity, and operability on moving platforms distinguishes the approach from both commercial microwave clocks and higher performing laboratory optical clocks. It compares favorably to active hydrogen masers in terms of long-term holdover while outperforming maser phase noise and instability at short timescales. To deliver peak performance, masers typically operate in large (~1,000 L) environmental chambers that carefully regulate the temperature and humidity, limiting their use to the laboratory. Conversely, no special measures were taken to control the operating environment of the iodine clock at both NIST and throughout the RIMPAC underway. Similar to cesium beam clocks, the 3U rackmount form factor lends itself to usage outside of the laboratory.

To our knowledge, these clocks are the highest performing sea-based clocks to date. The integration, packaging, and environmental robustness required to achieve such operation is a significant technological step toward widespread adoption of optical timekeeping. Since these field demonstrations, further advancement in the performance and SWaP of the rackmount clocks has been accomplished in our next-generation system [37] [38], including decreasing the short-term instability to $2\times10^{-14}/\sqrt{\tau}$, lowering the overall system SWaP to 30 L, 20 kg, and 70 W, and eliminating the external power supply.

## Acknowledgements

Andrew Dowd and Akash Rakholia supported electronics and code development. Micah Ledbetter helped to develop second-generation systems used for testing, and Eric Oelker carried out magnetic field evaluations. Jonathan Kohler and Matthew Cashen supported RIMPAC planning and operation and provided ship motion data collected during the underway. Leo

Hollberg provided useful feedback on iodine cells and spectroscopy. Jeff Sherman set up the maser link and provided UTC(NIST) ensemble data, and Laura Sinclair and Nathan Newbury supported the NIST clock measurements. Tommy Willis at the Office of Naval Research (ONR) coordinated the US effort for the Alternative PNT Challenge at RIMPAC under The Technical Cooperation Program (TTCP). Special thanks to the US Navy (NIWC Atlantic, NSWC) and the crew of the HMNZS Aotearoa for their warm hospitality and support of the at-sea demonstration.

This research was developed with funding from Army Research Laboratory (ARL) contract W911NF19C0047 (EPIC), Defense Advanced Research Projects Agency (DARPA) contracts 140D0420C0001 (HIPPOS), HR00111990037 (DILL PICKLES), and HR001121C0175 (PRICELESS), and Naval Air Systems Command (NAVAIR) contract N6833520C0116 (VIPER). The views, opinions and/or findings expressed are those of the authors and should not be interpreted as representing the official views or policies of the Department of Defense or the U.S. Government.

## Author Contributions

JDR, AC, WDL, GBP, ASK, DBS, and MMB designed and built the iodine spectrometer, laser system, and frequency comb. JDR, AC, WDL, GBP, and ASK operated the systems during field deployments. FR and MMB developed the control system and chassis. GES and JDR designed electronic subsystems for the physics packages. JPS and AC wrote firmware for the controllers, and WDL wrote the system level automation software. JRAS and MMB supervised the project.

## Methods

### Spectrometer

We interrogate the iodine $a_{10}$ hyperfine feature of the R(56) 32-0 transition via modulation transfer spectroscopy [17]. The vapor cell in which the spectroscopy is performed resides on an optical bench, and light is delivered to this bench via two fiber-coupled collimators. The pump and probe beams undergo multiple passes through the vapor cell in a racetrack configuration to provide adequate absorptive pathlength in a compact bench. Light is also sampled at various locations on the bench for implementing both RAM stabilization and power servos.

The PICKLES and EPIC spectrometers utilize a traditional vapor cell that includes a cold finger, which is temperature stabilized at a setpoint below 0 °C to maintain iodine pressure. The measured temperature-induced self-collisional frequency shift is ~-2 × $10^{-12}$/°C, in line with values from the literature [13]. Thus, the cold finger must be temperature stabilized at mK-levels to support a ~$10^{-15}$ flicker floor. This is accomplished with a custom-designed oven that surrounds the cold finger and has been able to maintain out-of-loop fluctuations < 1mK for time periods exceeding 10 days. The VIPER spectrometer employs a starved iodine cell, which carries a fixed quantity of gaseous iodine corresponding to the vapor pressure at the chosen fill temperature [39] [40]. This cell type eliminates the cold finger and removes the temperature dependence of the vapor density. As a result, the temperature coefficient is reduced by at least 10×, simplifying the stabilization requirement.

All optical elements routing the pump and probe beams on the optical bench are actively aligned and bonded during the build process. The result is an alignment-free subsystem. The bench, containing the routing optics and vapor cell, is enclosed in a thermal shield, which is also temperature stabilized. Two of the spectrometers (PICKLES and EPIC) are packaged into 2.5 L modules, which are surrounded by a single-layer magnetic shield. VIPER uses a smaller vapor cell and has relaxed thermal control requirements, resulting in a 1 L spectrometer module meant to target more modest frequency instability metrics. VIPER also does not include a magnetic shield. Light is introduced into each spectrometer with the two fibers, and all relevant error signals exit on electrical cables.

To understand the shielding requirements, the magnetic field sensitivity of the clock transition was explored, and a second-order Zeeman shift of magnitude $<10^{-15}/G^2$ was observed, which is $10^8\times$ smaller than for the 6.8 GHz rubidium microwave transition [41] and $10^4\times$ smaller than for the rubidium two-photon transition [25]. A linear Zeeman shift $< 10^{-14}/G$ is also observed for a field direction parallel to the laser beam propagation, which arises from an imperfect linear polarization of the pump beam. The singlet character of the molecular iodine clock transition provides an advantage compared to the relatively large alkali magnetic sensitivity.

**Laser System**

The iodine clock transition is compatible with industry-developed 1064 nm laser technology. The entire laser system is fiberized, permitting the use of telecom-packaged components. The laser system uses a low-noise, 1064 nm source and a portion of the 1064 nm light is passed to the frequency comb. The remaining infrared light is frequency-doubled and then split to create the pump and probe beams. The pump beam is frequency offset from the probe beam with an acousto-optic modulator (AOM) to avoid a coherent background signal due to spurious reflections [42]. In addition to the frequency offset, the AOM is used to frequency modulate (FM) the pump beam and add a controlled amplitude modulation (AM) to cancel residual amplitude modulation (RAM) detected at the spectrometer. This all-fiber laser system utilizes a single spool design for robustness and occupies a volume of 1 L.

The AC Stark shift has been measured to be $\sim 10^{-12}$/mW, requiring a power servo with relatively modest $\sim$1,000 ppm stability to reduce the contribution to the $\sim 10^{-15}$ level. This differs from clocks utilizing a two-photon optical clock transition, which typically demand at least an order of magnitude tighter control over power variations [25] [43].

The frequency shift with respect to residual amplitude modulation (RAM) has been measured to be $\sim 10^{-15}$/ppm, which is consistent with the $\sim$1 MHz clock transition linewidth (including pressure and power-broadening). Stabilizing the RAM to ppm-levels requires state-of-the-art control but is achievable [44]. Furthermore, frequency shifts from beam pointing errors, modulation amplitude dependence, de-modulation phase, and electronic baseline stability were quantified, and appropriate measures were implemented to minimize their contribution.

**Frequency Comb**

The all-fiber frequency comb utilizes erbium fiber technology and is related to the design described in Ref. [45]. The comb oscillator is centered at 1560 nm and operates at a repetition rate of 200 MHz. The pulse train output from this oscillator is amplified and broadened to both self-reference the comb and provide an extension to the clock subharmonic wavelength at 1064 nm. The derived CEO linewidth is ~250kHz FWHM with a beatnote SNR of ~45dB (300kHz RBW). Conversely, the optical beatnote has a SNR of ~40dB (300kHz RBW). The CEO beat frequency is stabilized by tuning the pump current, and the optical beatnote is stabilized with a high-bandwidth PZT actuator ($f_{3dB}$ bandwidth ~200kHz) that is incorporated into the oscillator. Measurement of the timing jitter between two identical frequency combs reveals that the combs can support a clock readout of $4 \times 10^{-17}$ at 1 s of averaging. The comb optics package occupies 0.5 L. Finally, conversion of the clock optical frequency to an RF output is accomplished with a microwave divider that converts the repetition rate to outputs at 100 MHz, 10 MHz, and 1 PPS.

**Clock Bring-up and Operation**

A software control system was developed that autonomously brings the system from an off-state to fully locked. Following power-up, the control system sets and monitors the temperature of all critical subcomponents. Once the system has reached target temperatures, an algorithm identifies the correct hyperfine manifold, and centers the laser on the appropriate hyperfine transition to lock the laser. The control system also engages the RAM and power servos for the pump and probe beams. The automated routine then locks the carrier envelope offset frequency of the comb to a pre-calibrated frequency and identifies the correct comb tooth for the optical lock prior to stabilizing the comb to the clock laser. Once the bring-up sequence is complete, the controller continually monitors the system state for any faults and takes the appropriate action to restore the system to a fully locked state.

## Supplemental Material

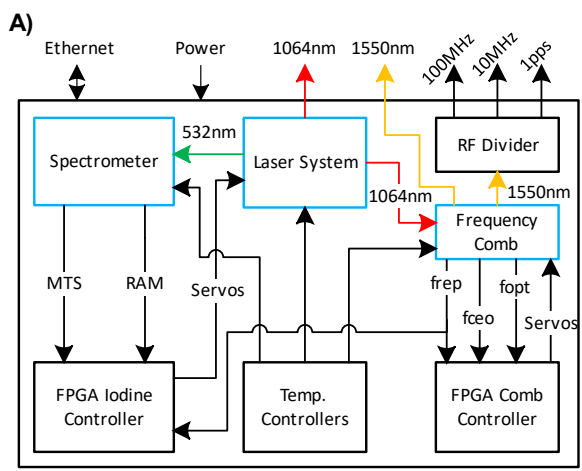
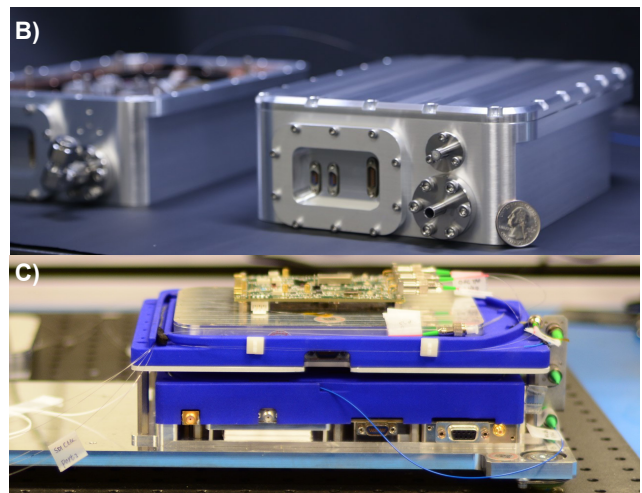

**Figure 4.** A) Block diagram of the clock illustrating the physics package subsystems and the control system interfaces within the chassis. The clock chassis includes a custom (B) spectrometer, (C) laser system and fiber frequency comb optics packages (bottom and top of image, respectively).

A block diagram for the optical clock system illustrating the physics package and electronics control system is shown in Figure 4A. Figure 4B illustrates the 2.5 L spectrometer housings that are used for the PICKLES and EPIC systems. The 532 nm laser system (bottom) and 1550 nm frequency comb (top) are seen in Figure 4C. The combination of the laser system and frequency comb optics packages occupy a total volume of 1.5 L.

Figure 5A is a photograph of the PICKLES and EPIC systems during the NIST measurement campaign. Both systems were placed on an optical table without any additional environmental shielding. Each of these systems has its own 1U rackmount power supply (not visible) and control laptop. The 10 MHz microwave output from each clock was compared against a 5 MHz tone from the NIST maser ST05 with a Microchip 53100A phase noise analyzer in a three-cornered hat configuration (Figure 5B and seen in the center of the photograph). The 1064-nm optical beatnote between the PICKLES and EPIC systems was also monitored for cross-validation. Once installed, the clocks and measurement apparatus operated autonomously. The system telemetry was remotely monitored without intervention from our California office.

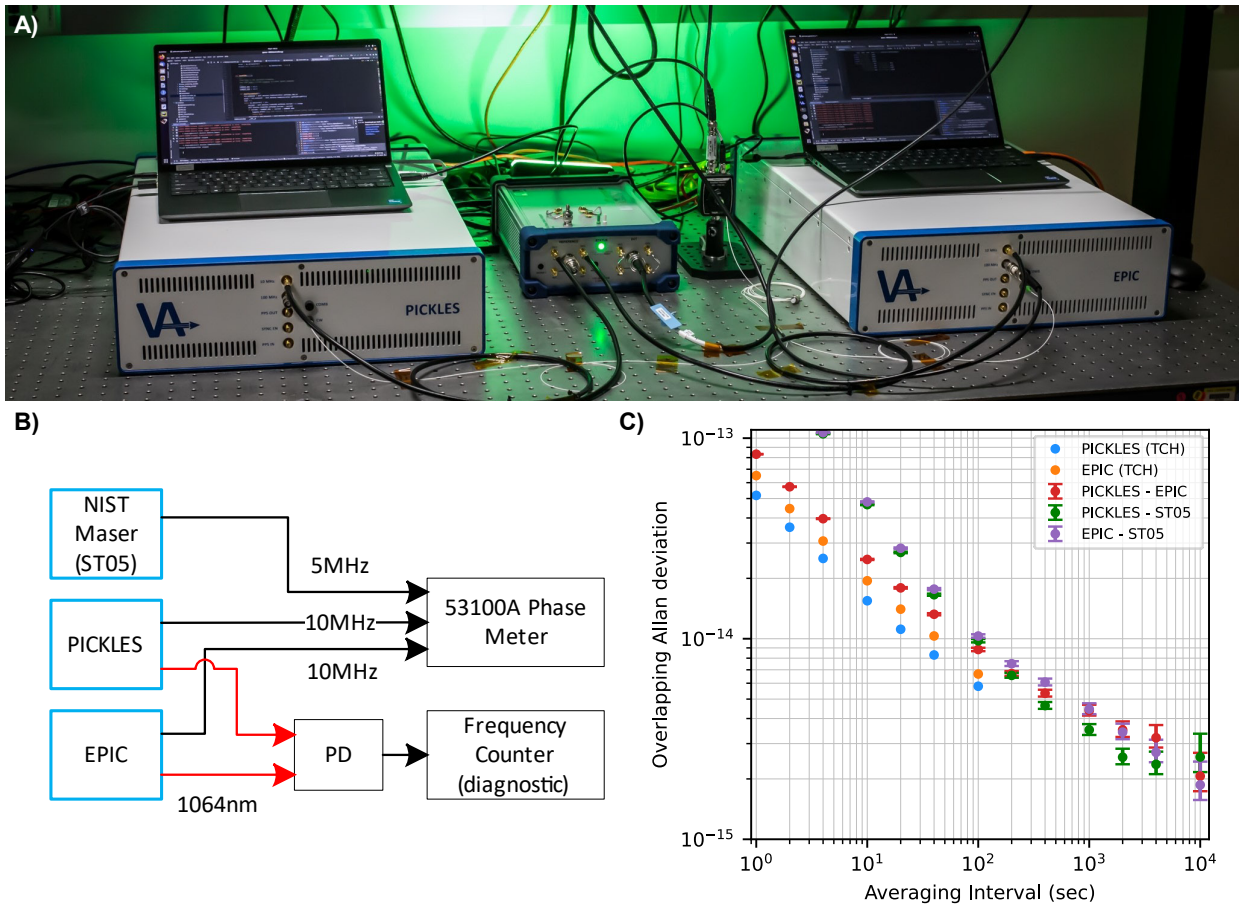

Figure 5. Clock characterization at NIST. A) Two independent 3U, 19-inch rackmount optical clocks. Front panel outputs are 10 MHz, 100 MHz, and 1 PPS signals in addition to stabilized 1064 nm clock and 1550 nm comb light. B) Block diagram for the measurement of two iodine clocks versus the ST05 maser at NIST Boulder. The 10 MHz outputs from each clock were compared to the 5 MHz output from ST05; diagnostics were included to monitor the 1064 nm beatnote in parallel. C) Raw instability results for the three pairwise microwave comparisons over the course of 24 hours. These 1-day subsets show both clock instabilities at ~$2\times10^{-15}$ after $10^4$ seconds of averaging. Three cornered hat (TCH) extraction of the individual clock instabilities show that the two Iodine clocks operate at $5\times10^{-14}/\sqrt{\tau}$ and $6\times10^{-14}/\sqrt{\tau}$.

A representative measurement is shown in Figure 5C. The raw PICKLES-EPIC instability exhibits short-term performance of $8\times10^{-14}/\sqrt{\tau}$. Data collection for ~ 1 day permits reliable estimation of the clock performance at 10,000 s, and noise averaging down to ~$2\times10^{-15}$ is observed. Decorrelation of the three clocks using the three-cornered hat methodology reveals PICKLES and EPIC short-term instabilities of $5\times10^{-14}/\sqrt{\tau}$ and $6\times10^{-14}/\sqrt{\tau}$, respectively, outperforming the short-term performance of the ST05 maser. Excellent agreement is observed between the microwave and optical beatnotes (not shown).

The microwave phase noise between PICKLES and EPIC (seen in Figure 1B) was measured with the 53100A phase noise analyzer, and the phase noise for a single iodine clock was inferred by subtracting 3dB from the measured result. In particular, the phase noise at a 1 Hz offset for the 10 MHz (100 MHz) output is -124 (-105) dBc/Hz before decreasing to a white phase noise floor at -170 (-155) dBc/Hz. More specifically, the measured phase noise $\mathcal{L}(f)$ at both frequencies may be fit to the form: $\mathcal{L}(f) = \frac{1}{2} \cdot \sum_{n=-2}^{0} 10^{b_n} \cdot f^n$, where the coefficients $b = [b_{-2}, b_{-1}, b_0]$ correspond to white frequency, flicker phase, and white phase noise, respectively [46]. The observed phase noise coefficients at 10 MHz are $b^{10MHz} = [-12.2, -13.2, -16.7]$, whereas those at 100 MHz are $b^{100MHz} = [-10.2, -11.9, -15.2]$. The phase noise at both frequencies is lower than commercial atomic-disciplined oscillators, highlighting the benefits of optical frequency division where the fractional noise of the iodine-stabilized laser is transferred to the RF comb repetition rate.

Similarly, the optical phase noise at 1064 nm displayed in Figure 1B is measured between the PICKLES system and a higher performing iodine clock. The phase noise at a 1Hz offset is 23 dBc/Hz and the measured phase noise $\mathcal{L}(f)$ follows the form described above with the coefficients $b^{1064nm} = [26,0,0]$ down to the servo bump, which is visible at ~10 kHz.

The experimental block diagram describing the measurements during RIMPAC 2022 is shown in Figure 6A. The 19-inch rack containing the PICKLES, EPIC, and VIPER systems includes a 53100A phase noise analyzer to compare the 100 MHz tones derived from each clock's frequency comb against the other two in a three-cornered hat configuration. As with the NIST setup, the three pairwise optical beatnotes between the three systems was also collected for cross-validation (not shown).

Frequency time-traces for the three pairwise comparisons at 100 MHz are shown in Figure 6B. For consistency, all the collected data is shown following the VIPER external power supply restart. As described in the main text, the VIPER physics package is an earlier design with relaxed performance goals. A diurnal temperature variation is evident in the VIPER system. Nonetheless, VIPER outperforms clocks conventionally used at sea.

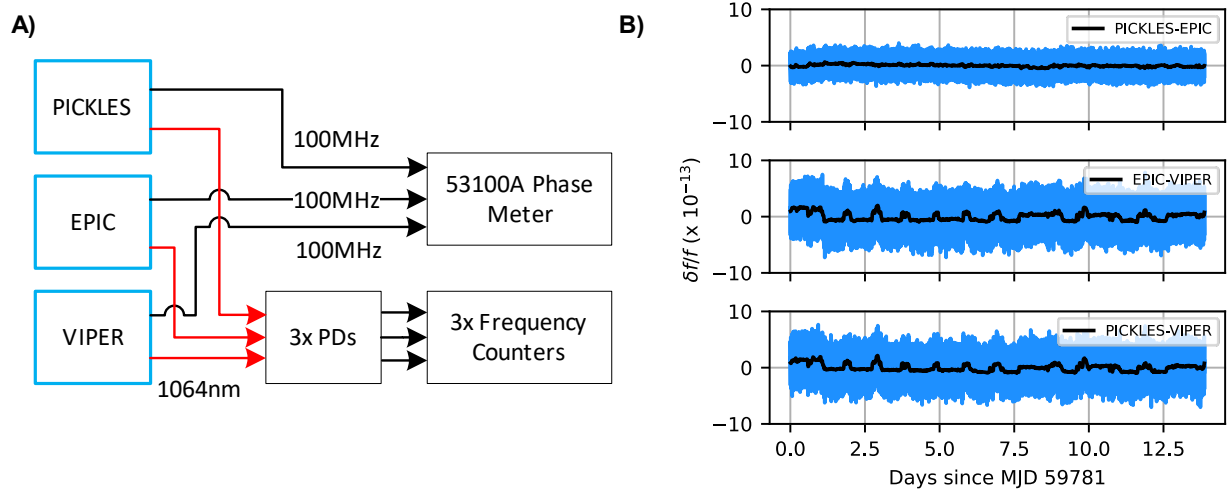

**Figure 6. Clock comparison at sea.** A) Block diagram for the measurement of three iodine clocks during RIMPAC 2022. The 100 MHz output from each clock was input to a Microsemi 53100A phase noise analyzer in a three-cornered hat configuration. The three pairwise optical beatnotes at 1064 nm were also collected in parallel. B) Time series for the three pairwise comparisons at 100 MHz over fourteen days at sea. The blue trace in each panel is the fractional frequency noise with a gate time of 1 s. The black trace is a 1,000 s moving average.

# References

[1] J. C. Koelemeij, H. Dun, C. E. Diouf, E. F. Dierikx, G. J. Janssen and C. C. Tiberius, "A hybrid optical–wireless network for decimetre-level terrestrial positioning," *Nature,* vol. 611, no. 7936, p. 473, 2022.

[2] G. Marra and et al., "Optical interferometry–based array of seafloor environmental sensors using a transoceanic submarine cable," *Science,* vol. 376, no. 6595, p. 874, 2022.

[3] C. Clivati and et al., "Common-clock very long baseline interferometry using a coherent optical fiber link," *Optica,* vol. 7, no. 8, p. 1031, 2020.

[4] C. Clivati and et al., "Coherent phase transfer for real-world twin-field quantum key distribution," *Nature communications,* vol. 13, no. 1, p. 157, 2022.

[5] S. Wang and et al., "Twin-field quantum key distribution over 830-km fibre," *Nature photonics,* vol. 16, no. 2, p. 154, 2022.

[6] BACON Collaboration, "Frequency ratio measurements at 18-digit accuracy using an optical clock network," *Nature,* vol. 7851, pp. 564-569, 2021.

[7] A. D. Ludlow, M. M. Boyd, J. Ye, E. Peik and P. O. Schmidt, "Optical atomic clocks," *Reviews of Modern Physics,* vol. 87, no. 2, p. 637, 2015.

[8] J. Grotti and et al., "Geodesy and metrology with a transportable optical clock," *Nature Physics,* vol. 14, no. 5, p. 437, 2018.

[9] M. Takamoto, I. Ushijima, N. Ohmae, T. Yahagi, K. Kokado, H. Shinkai and H. Katori, "Test of general relativity by a pair of transportable optical lattice clocks," *Nature Photonics,* vol. 14, no. 7, p. 411, 2020.

[10] B. L. S. Marlow and D. R. Scherer, "A review of commercial and emerging atomic frequency standards," *IEEE Transactions on Ultrasonics, Ferroelectrics, and Frequency Control,* vol. 68, no. 6, pp. 2007-2022, 2021.

[11] M. Lombardi, "An evaluation of dependencies of critical infrastructure timing systems on the global positioning system (gps)," NIST Technical Note 1289, 2021.

[12] E. D. Caldwell, J. D. Deschenes, J. Ellis, W. C. Swann, B. K. Stuhl, H. Bergeron, N. R. Newbury and L. C. Sinclair, "Quantum-limited optical time transfer for future geosynchronous links," *Nature,* vol. 618, no. 7966, p. 721, 2023.

[13] M. L. Eickhoff and J. L. Hall, "Optical frequency standard at 532 nm," *IEEE transactions on instrumentation and measurement,* vol. 44, no. 2, pp. 155-158, 1995.

[14] A. Arie and R. L. Byer, "Laser heterodyne spectroscopy of 127 I2 hyperfine structure near 532 nm," *JOSA B,* vol. 10, no. 11, pp. 1990-1997, 1993.

[15] P. A. Jungner, S. Swartz, M. Eickhoff, J. Ye, J. L. Hall and S. Waltman, "Absolute frequency of the molecular iodine transition R (56) 32-0 near 532 nm.," *IEEE transactions on instrumentation and measurement,* vol. 44, no. 2, pp. 151-154, 1995.

[16] T. W. Hänsch, M. D. Levenson and A. L. Schawlow, "Complete hyperfine structure of a molecular iodine line," *Physical Review Letters,* vol. 26, no. 16, p. 946, 1971.

[17] J. Ye, L. Robertsson, S. Picard, L. S. Ma and J. L. Hall, "Absolute frequency atlas of molecular I2 lines at 532 nm," *IEEE Transactions on Instrumentation and Measurement,* vol. 48, no. 2, p. 544, 1999.

[18] F. Riehle, P. Gill, F. Arias and L. Robertsson, "The CIPM list of recommended frequency standard values: guidelines and procedures," *Metrologia,* vol. 55, no. 2, p. 188, 2018.

[19] J. Ye, L. S. Ma and J. L. Hall, "Molecular iodine clock," *Physical review letters,* vol. 87, no. 27, p. 270801, 2001.

[20] K. Döringshoff, T. Schuldt, E. V. Kovalchuk, J. Stühler, C. Braxmaier and A. Peters, "A flight-like absolute optical frequency reference based on iodine for laser systems at 1064 nm," *Applied physics B,* vol. 123, pp. 1-8, 2017.

[21] T. Schuldt, K. Döringshoff, E. V. Kovalchuk, A. Keetman, J. Pahl, A. Peters and C. Braxmaier, "Development of a compact optical absolute frequency reference for space with 10–15 instability," *Applied optics,* vol. 56, no. 4, pp. 1101-1106, 2017.

[22] K. Döringshoff and et al., "Iodine frequency reference on a sounding rocket," *Physical Review Applied,* vol. 11, no. 5, p. 054068, 2019.

[23] A. Mehlman and et al., "Iodine based reference laser for ground tests of LISA payload," in *International Conference on Space Optics 2022*, Dubrovnik, Croatia, 2022.

[24] C. Perrella, P. S. Light, J. D. Anstie, F. N. Baynes, R. T. White and A. N. Luiten, "Dichroic two-photon rubidium frequency standard," *Physical Review Applied,* vol. 12, no. 5, p. 054063, 2019.

[25] K. Martin, G. Phelps, N. Lemke, M. Bigelow, B. Stuhl, M. Wojcik, M. Holt, I. Coddington, M. Bishop and J. Burke, "Compact optical atomic clock based on a two-photon transition in rubidium," *Physical Review Applied,* vol. 9, no. 1, p. 014019, 2018.

[26] Z. L. Newman and et al., "High-performance, compact optical standard," *Optics Letters,* vol. 46, no. 18, p. 4702, 2021.

[27] J. A. Sherman, L. Arissian, R. C. Brown, M. J. Deutch, E. A. Donley, V. Gerginov, J. Levine, G. K. Nelson, A. N. Novick, B. R. Patla and T. E. Parker, "A resilient architecture for the realization and distribution of coordinated universal time to critical infrastructure systems in the United States," *NIST Technical Note 2187,* 2021.

[28] M. Epstein, T. Dass, J. Rajan and P. Gilmour, "Long-term clock behavior of GPS IIR satellites," *Proceedings of the 39th Annual Precise Time and Time Interval Meeting,* pp. 59-78, 2007.

[29] J. C. Camparo, J. O. Hagerman and T. A. McClelland, "Long-term behavior of rubidium clocks in space," *2012 European Frequency and Time Forum,* pp. 501-508, 2012.

[30] T. E. Parker, "Environmental factors and hydrogen maser frequency stability," *IEEE transactions on ultrasonics, ferroelectrics, and frequency control,* vol. 46, no. 3, p. 745, 1999.

[31] J. Sherman, *private communication,* 2023.

[32] A. D. Ludlow, X. Huang, M. Notcutt, T. Zanon-Willette, S. M. Foreman, M. M. Boyd, S. Blatt and J. Ye, "Compact, thermal-noise-limited optical cavity for diode laser stabilization at 1× 10-15," *Optics letters,* vol. 32, no. 6, p. 641, 2007.

[33] J. Kohler and et al., "Development of a STRAP-DOWN, Absolute Atomic Gravimeter for MAPMatching Navigation," in *IEEE International Symposium on Inertial Sensors & Systems*, Kaua'i, Hawaii, 2023.

[34] A. Hilton and et al., "Demonstration of a Field-Deployable Ytterbium Cell Clock," in *IEEE IFCS-EFTC 2023*, Toyama, Japan, 2023.

[35] K. W. Martin, R. Beard and J. Elgin, "Testing an Optical Atomic Clock in the Field," in *Joint Navigation Conference*, San Diego, California, 2023.

[36] https://www.ngdc.noaa.gov/geomag/geomag.shtml.

[37] J. Roslund, A. Cingoz, A. Kowligy, W. Lunden, G. Partridge, F. Roller, D. Sheredy, G. Skulason, J. Song, J. Abo-Shaeer and M. Boyd, "Deployable Optical Atomic Clocks," in *Precise Time and Time Interval Meeting*, 2023.

[38] A. Cingoz, A. Kowligy, J. Roslund, J. Abo-Shaeer, M. Boyd, W. Lunden, F. Roller, E. Caldwell, N. Newbury, L. Sinclair, B. Stuhl and W. Swann, "Integrated Optical Transceivers for Frequency Comb-Based Time Transfer over Free Space," in *Precise Time and Time Interval Meeting*, 2023.

[39] J. Hrabina, M. Šarbort, O. Acef, F. Du Burck, N. Chiodo, M. Holá, O. Číp and J. Lazar, "Spectral properties of molecular iodine in absorption cells filled to specified saturation pressure," *Applied optics,* vol. 53, no. 31, pp. 7435-7441, 2014.

[40] L. Hollberg, "Surprisingly good frequency stability from tiny green lasers and iodine molecules," in *Photonics West*, San Francisco, California, 2018.

[41] J. Vanier and C. Audoin, The quantum physics of atomic frequency standards, CRC Press, 1989.

[42] J. J. Snyder, R. K. Raj, D. Bloch and M. Ducloy, "High-sensitivity nonlinear spectroscopy using a frequency-offset pump," *Optics Letters,* vol. 5, no. 4, pp. 163-165, 1980.

[43] K. Martin, B. Stuhl, J. Eugenio, M. Safronova, G. Phelps, J. Burke and N. Lemke, "Frequency shifts due to Stark effects on a rubidium two-photon transition," *Physical Review A,* vol. 100, no. 2, p. 023417, 2019.

[44] W. Zhang, M. Martin, C. Benko, J. Hall, J. Ye, C. Hagemann, T. Legero, U. Sterr, F. Riehle, G. Cole and M. Aspelmeyer, "Reduction of residual amplitude modulation to 1× 10-6 for frequency modulation and laser stabilization," *Optics letters,* vol. 39, no. 7, pp. 1980-1983, 2014.

[45] L. C. Sinclair, J.-D. Deschênes, L. Sonderhouse, W. C. Swann, I. H. Khader, E. Baumann, N. R. Newbury and I. Coddington, "A compact optically coherent fiber frequency comb," *Review of Scientific Instruments,* p. 081301, 2015.

[46] F. Riehle, Frequency standards: basics and applications, John Wiley & Sons, 2006.

[47] Petty Officer 3rd Class Taylor Bacon, Artist, *USCGC Midgett conducts refueling at sea [Image 2 of 3].* [Art]. 2022.